\RequirePackage{setspace}
\documentclass[sigconf,natbib=true]{acmart}
\usepackage{stfloats}
\usepackage{multirow}
\usepackage{subfigure} 
\usepackage{makecell}
\usepackage{caption}
\usepackage{subfigure}
\usepackage{arydshln}
\usepackage{booktabs}
\usepackage{multirow}
\usepackage{tablefootnote}
\usepackage{setspace}
\usepackage{hyperref}
\AtBeginDocument{%
  \providecommand\BibTeX{{%
    \normalfont B\kern-0.5em{\scshape i\kern-0.25em b}\kern-0.8em\TeX}}}

\copyrightyear{2022} \acmYear{2022} \setcopyright{rightsretained} \acmConference[CIKM '22]{Proceedings of the 31st ACM International Conference on Information and Knowledge Management}{October 17--21, 2022}{Atlanta, GA, USA} \acmBooktitle{Proceedings of the 31st ACM International Conference on Information and Knowledge Management (CIKM '22), October 17--21, 2022, Atlanta, GA, USA} \acmDOI{10.1145/3511808.3557388} \acmISBN{978-1-4503-9236-5/22/10}
\begin{document}

%\title{General Text Matching Model for Multiple Tasks Based on Prompt Learning}
\title{Match-Prompt: Improving Multi-task Generalization Ability for Neural Text Matching via Prompt Learning}

\author{Shicheng Xu}
\orcid{0000-0001-7157-3410}
\affiliation{%
   \institution{Data Intelligence System Research Center, Institute of Computing Technology, CAS \\ University of Chinese Academy of Sciences}
  \city{Beijing}
  \country{China}}
\email{xschit@163.com}

\author{Liang Pang}
%\blfootnote{Corresponding author}
\authornote{Corresponding author}
\affiliation{%
   \institution{Data Intelligence System Research Center, Institute of Computing Technology, CAS}
  \city{Beijing}
  \country{China}}
\email{pangliang@ict.ac.cn}

\author{Huawei Shen}
\affiliation{%
   \institution{Data Intelligence System Research Center, Institute of Computing Technology, CAS \\ University of Chinese Academy of Sciences}
  \city{Beijing}
  \country{China}}
\email{shenhuawei@ict.ac.cn}

\author{Xueqi Cheng}
\affiliation{%
   \institution{CAS Key Lab of Network Data Science and Technology, Institute of Computing Technology, CAS \\ University of Chinese Academy of Sciences}
  \city{Beijing}
  \country{China}}
\email{cxq@ict.ac.cn}

\begin{abstract}
Text matching is a fundamental technique in both information retrieval and natural language processing. Text matching tasks share the same paradigm that determines the relationship between two given texts.  
The relationships vary from task to task, e.g.~relevance in document retrieval, semantic alignment in paraphrase identification and answerable judgment in question answering.
However, the essential signals for text matching remain in a finite scope, i.e.~exact matching, semantic matching, and inference matching. Ideally, a good text matching model can learn to capture and aggregate these signals for different matching tasks to achieve competitive performance, while recent state-of-the-art text matching models, e.g.~Pre-trained Language Models (PLMs), are hard to generalize. It is because the end-to-end supervised learning on task-specific dataset makes model overemphasize the data sample bias and task-specific signals instead of the essential matching signals, which ruins the generalization of model to different tasks. 
To overcome this problem, we adopt a specialization-generalization training strategy and refer to it as Match-Prompt. In specialization stage, descriptions of different matching tasks are mapped to only a few prompt tokens. 
In generalization stage, text matching model explores the essential matching signals by being trained on diverse multiple matching tasks. 
High diverse matching tasks avoid model fitting the data sample bias on a specific task, so that model can focus on learning the essential matching signals. Meanwhile, the prompt tokens obtained in the first step are added to the corresponding tasks to help the model distinguish different task-specific matching signals, as well as to form the basis prompt tokens for a new matching task.
In this paper, we consider five common text matching tasks including document retrieval, open-domain question answering, retrieval-based dialogue, paraphrase identification, and natural language inference.
Experimental results on eighteen public datasets show that Match-Prompt can improve multi-task generalization capability of PLMs in text matching and yield better in-domain multi-task, out-of-domain multi-task and new task adaptation performance than multi-task and task-specific models trained by previous fine-tuning paradigm.
\end{abstract}





\maketitle

\section{Introduction} \label{introduction}
Text matching is an important technique that can find accurate information from the huge amount of resources and plays a fundamental role in many downstream tasks, such as Document Retrieval (DR)~\cite{adhoc}, Open-domain Question Answering (QA)~\cite{open-domain-qa}, Retrieval-based Dialogue (RD)~\cite{ac}, Paraphrase Identification (PI)~\cite{pi}, and Natural Language Inference (NLI)~\cite{nli}. 
Traditional text matching methods focus on measuring the word-to-word exact matching between two texts, such as TF-IDF~\cite{tf-idf} and BM25~\cite{bm25}. Such methods follow the heuristics in information retrieval~\cite{ir} and can generalize to different tasks by adjusting few parameters (e.g. $k_1$ and $b$).
To introduce semantic and other complex relationships into text matching models, deep learning is applied to project texts into a joint and semantic space, namely neural text matching. Especially, PLMs like BERT~\cite{bert} have brought great improvement, but the multi-task generalization of them has not been well explored. In general, most deep models can only achieve good performance when trained and evaluated on a single specific dataset, while have poor generalization ability to different domains and tasks~\cite{linguistic}.

To deal with the above problem of text matching models, we begin with the definition and evaluation of multi-task generalization. In this paper, the multi-task generalization of a text matching model can be divided into three levels: 1) in-domain multi-task, model has access to different datasets in multiple matching tasks, e.g. QA, RD, PI, NLI, and DR, and evaluates on each given datasets. It reflects the ability to learn from diverse matching tasks and apply it to current domain of each task. 2) out-of-domain multi-task, model can access the same datasets as the in-domain multi-task level, but is evaluated on the out-of-domain unseen datasets. It reflects the ability to apply learned knowledge to a new domain, e.g. from financial QA to medical QA. 3) new task adaptation, making a task invisible to the model, e.g. QA, and training model on other tasks, e.g. RD, PI, NLI, DR. Given few-shot examples in unseen task to adapt the model to this new task. It reflects the ability of new task generalization and this is the hardest generalization level for matching models. 

\begin{figure}
  \centering
  \includegraphics[width=\linewidth]{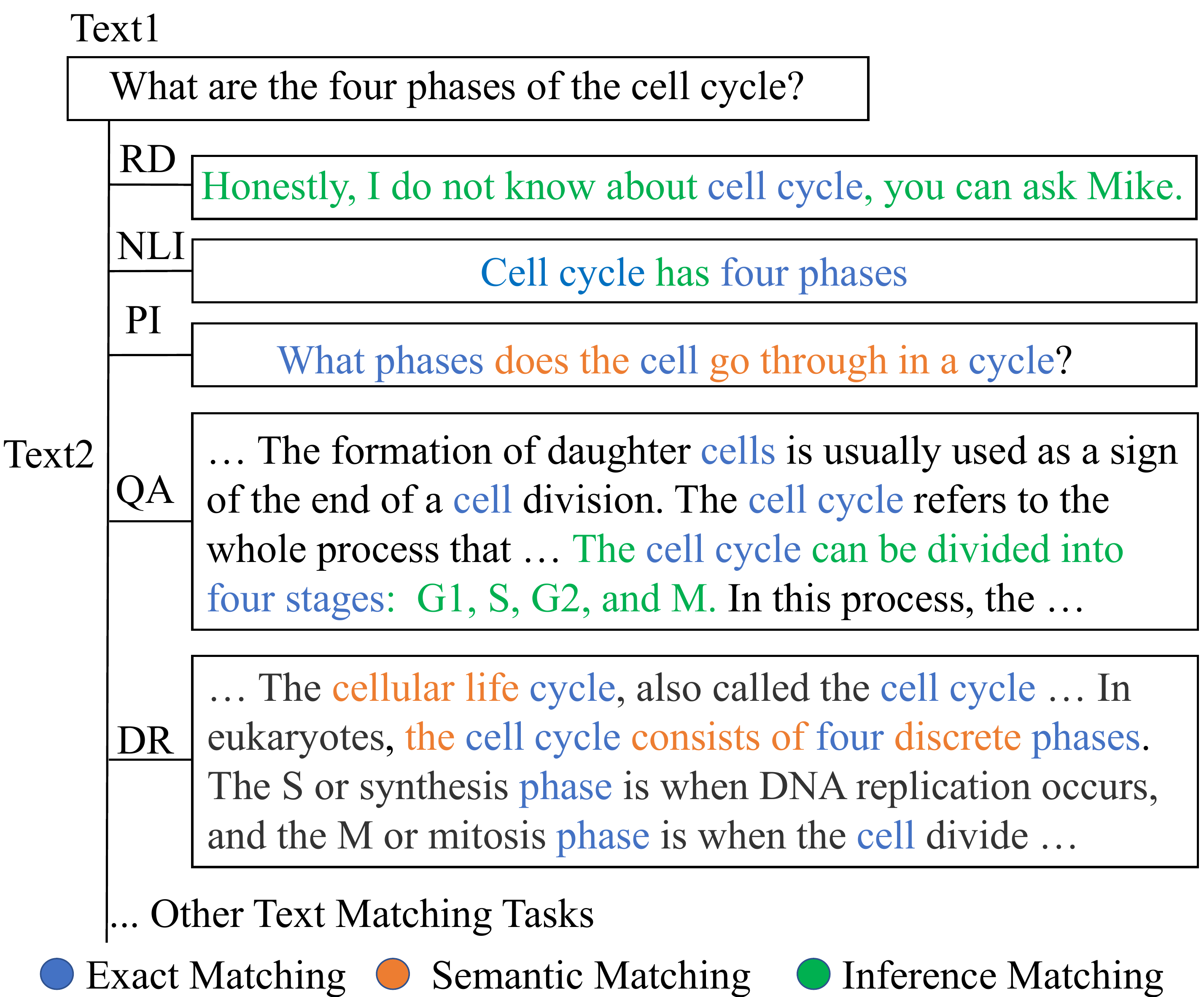}
  \caption{Matching texts and signals for {\itshape "What are the four phases of the cell cycle?"} in different tasks.}
  \label{figure:example}
\end{figure}

The main challenges to obtaining a good generalized text matching model lay in two folds, 1) how to learn generalized matching signals? 2) how to apply a multi-task model to different matching tasks?
From the learning perspective, the end-to-end supervised learning on task-specific datasets makes the model overemphasize the data sample bias and task-specific signals instead of the essential matching signals. For example in Figure~\ref{figure:example}, learning only the PI task causes the model to over-focus on the exact match signals, such as `cell', `cycle', as well as question mark which make it hard to generalize. 
However, there are matching signals that are shared across tasks, and in this paper, we classify matching signals into three categories, namely exact matching signals, semantic matching signals and inference matching signals. 
Exact matching signals focus on the word overlapping. It is shown that exact matching signals are important in all text matching tasks, for example, query term importance in DR~\cite{drmm, deeprank}, lexical overlap features in PI~\cite{pipqr} and bigram counts in QA~\cite{open-domain-qa}. 
Semantic matching signals measure semantic alignment. Many studies prove that introduce semantic information into DR~\cite{dssm} and PI~\cite{matchpyramid} can improve the performance, especially for QA that dense retrieval~\cite{dense-retrival} is better than sparse retrieval, e.g. DPR~\cite{dpr} and ORQA~\cite{orqa}. 
Inference matching signals infer implication relationship. NLI, QA and RD have to infer implication relation between two texts. For example, RD needs to confirm the reply is a logical consequence of the post and dialogue history, and QA needs to pick out the answer that the question hardly contains.
From the applying perspective, the difference between text matching tasks is the different fusion and utilization of these matching signals.
Thus, if one matching model can capture the matching signals shared across tasks and combine them according to a specific matching task, its multi-task generalization ability will be improved. 

Inspired by prompt learning, we propose a multi-task text matching model that exhibits strong multi-task generalization ability at all three levels. Prompt learning is a new PLMs tuning strategy that transforms various NLP tasks into the form of [MASK] prediction, which is more in line with the pre-training tasks of PLMs. In our method, we adopt a specialization-generalization training strategy and refer to it as Match-Prompt. In the specialization stage, prompts of each specific text matching task are obtained by specially designed prompt engineering that maps the description of different text matching tasks into only a few prompt tokens. In the generalization stage, the PLM is trained on mixed dataset consisting of different text matching tasks, and the prompt tokens from the first stage are used as task-discriminative tokens. Diversifying text matching tasks help the model explore the essential matching signals instead of overemphasizing the data sample bias and task-specific signals, to exploit shared information from tasks and address the first challenge. The prompt tokens of each task are added to the corresponding tasks to distinguish different tasks to answer the second challenge. Besides, unlike traditional prompt learning, it is shown that the learned task-specific prompts obtain the information to combine the matching signals, in particular, we can construct a proper prompt for new task by simply combining other learned prompts.
 
Experimental results on eighteen public datasets show that our method yields better in-domain multi-task, out-of-domain multi-task and new task adaptation performance compared to the baseline fine-tuning paradigm. The results also indicate that Match-Prompt has a stronger ability to distinguish tasks and utilize essential matching signals shared by multiple tasks. Both of them are beneficial for improving the multi-task generalization ability.

To sum up, our contributions are as follows:
\begin{itemize}
\item{} We collect eighteen datasets from diverse text matching tasks, which can serve as a benchmark for the multi-task generalization ability of text matching models.
\item{} We propose a specialization-generalization training strategy based on prompt learning to disentangle general matching signal learning and specific task combination. 
\item{} We divide multi-task generalization into three levels and Match-prompt can enhance the multi-task generalization ability of BERT in text matching at these three levels. Code and datasets are released at \url{https://github.com/xsc1234/Match-Prompt}.
\end{itemize}

\section{Related Work}
In this section, we review the previous studies on text matching, mixed training for multiple tasks and prompt learning. 

\subsection{Text Matching}

Traditional text matching methods, such as TF-IDF~\cite{tf-idf} and BM25~\cite{bm25}, represent texts as sparse vectors, each element in vectors corresponds to a term. They pay attention to the exact matching between two texts at term-level and can get good generalization performance. However, these methods suffer from the lexical gap and cannot capture the semantic information of the text. 

Neural text matching models have brought great improvement with the help of deep neural networks and the training on a lot of in-domain data. We introduce the related work of neural text matching from three aspects: exact matching, semantic matching and inference matching. In exact matching, DRMM~\cite{drmm} considers query item importance and is suitable for DR. DrQA~\cite{open-domain-qa} combines TF-IDF weighted bag-of-words vectors and bigram to represent the text. This method shows good performance in open-domain QA. In semantic matching, interaction-based text matching models such as DeepMatch~\cite{deepmatch}, ARC-\uppercase\expandafter{\romannumeral2}~\cite{arc2}, MatchPyramid~\cite{matchpyramid}, ESIM~\cite{esim} and Match-SRNN~\cite{SRNN} can describe the fine-grained semantic matching relationship between two texts. They are often used in short text matching tasks such as PI, NLI and RD. Models based on single semantic text representation such as DSSM~\cite{dssm}, CDSSM~\cite{cdssm}, and ARC-I~\cite{arc1} represent text as a dense vector to get the matching score. These models ensure retrieval efficiency and are often used in DR and QA. In inference matching that needs models to infer new information from the text, asymmetric neural text matching models such as DeepRank~\cite{deeprank} and RE2~\cite{re2} are suitable. More recently, PLMs have achieved state-of-the-art text matching performance, such as BERT~\cite{bert}. In this paper, we mainly discuss the cross-attention rather than the bi-encoder form of BERT because the former can achieve better matching effect. 

The previous text matching models are only suitable for one of the specific tasks according to their specifically designed structures and mechanisms. Even though PLMs can be applied in multiple text matching tasks, only task-specific models that are fine-tuned on specific tasks can achieve good performance. When using the traditional fine-tuning method to train PLMs on mixed datasets of multiple matching tasks, the performance of the model drops seriously~\cite{linguistic}. Our approach focuses on how to allow PLMs to distinguish different tasks among multiple tasks and fuse the shared information of each task during multi-task learning to obtain stronger multi-task generalization ability.

\subsection{Mixed Training for Multiple Tasks}

In order to enable a single model to complete multiple tasks, a common strategy is training the model on the mixed datasets of multiple tasks. A straightforward approach is directly mixing the datasets of each task without any task-specific marks and fine-tuning PLMs on mixed datasets. For example,  Alon et al.~\cite{multiqa} propose MultiQA. They train model on multiple datasets and find it leads to robust generalization in reading comprehension task. Jean et al.~\cite{multi-retrieval} propose a multi-task retrieval model for knowledge-intensive tasks. They use KILT~\cite{KILT} as the mixed datasets without any task-specific tokens. Some approaches add additional task-specific components containing parameters for each task. Liu et al.~\cite{MT-DNN} propose MT-DNN that uses shared transformer encoder to encode multiple NLU tasks and add task-specific layers for each task. There are also some methods that transform multiple tasks into a unified question answering format such as MQAN~\cite{MQAN}, UnifiedQA~\cite{unifiedqa} and T5~\cite{t5}. Li et al.~\cite{ie} apply this method to information extraction, Chai et al.~\cite{tc} and Puri et al.~\cite{zero-shot-tc} use it in text classification. 

The main tasks of the above methods are not text matching. In addition, some of these methods cannot fully utilize the differentiating marks, some introduce additional task-specific layers, and some need super-large-scale models. Compared with previous methods, Match-Prompt can improve the multi-task generalization capability of BERT in text matching and we do not need to add any task-specific layers when using the model for prediction.

\subsection{Prompt Learning}
Prompt learning is a novel tuning strategy for PLMs. It converts multiple NLP tasks into the form of [MASK] prediction using LM by adding template to the input texts. How to create the template is very important for prompt learning. On the whole, it can be divided into two categories, one is manual template engineering, the other is automated template learning~\cite{promptzongshu}. In manual template engineering, Petroni et al.~\cite{lama} design manual template to probe knowledge in LMs. Schick et al.~\cite{pet} proposes Pattern-Exploiting Training (PET), which combines few-shot learning with templates, and converts specific tasks into cloze tasks. Although creating templates manually can solve many tasks, it has some issues: 1) making templates requires a lot of knowledge and experience; 2) manually designed templates may not be globally optimal~\cite{prefix}. To address these problems, there are many automated template learning methods proposed such as token-based gradient searching~\cite{searching}, mining training corpus~\cite{mining}, Prefix Tuning~\cite{prefix}, and P-tuning~\cite{p-tunning}. Recently, there are some papers on pre-training or fine-tuning on multiple tasks, such as SPoT~\cite{spot}, PPT~\cite{ppt}, ExT5~\cite{ext5}, FLAN~\cite{instruction-learning}, ZeroPrompt~\cite{zeroprompt} and zero-shot for task generation~\cite{zero-generation}. Different from previous studies, we do not focus on how to improve the performance of prompt learning, but how to apply it to enhance generalization ability of text matching models. We also explore the relationships of task prompt tokens and find that a new matching task prompt can be constructed by linear combining other learned task prompt tokens.

\section{Match-Prompt} \label{methods}
In this section, we elaborate on the details of Match-Prompt and construct a multi-task text matching model.
\subsection{Overall Framework}

Match-Prompt is a two-stage training strategy including specialization and generalization. In specialization stage, the descriptions of each task are mapped to a few prompt tokens by prompt engineering. These tokens contain the information of each task and can be used as the differentiating marks of each task in multi-task learning. The second stage is generalization. Prompt tokens obtained in specialization stage are added to the input text of the corresponding specific task and BERT is trained on mixed datasets consisting of different tasks. In this stage, high diverse matching tasks avoid the model fitting the data sample bias on a specific task, so that model can focus on learning the common and essential matching signals. Besides, prompt tokens obtained in the specialization stage help the multi-task model better distinguish different tasks, which is beneficial for improving its multi-task generalization ability.

Compared with traditional fine-tuning paradigm, Match-Prompt offers several advantages. First, same as other prompt learning methods, Match-Prompt can make full use of the knowledge of PLMs and has better generalization ability ~\cite{pet,prefix,lama}. Second, Match-Prompt can provide better differentiation for each task in mixed training and predicting of the model. Third, the differentiating marks of each task are added to the input text in the form of prompt tokens, and the word at [MASK] is predicted by PLM, which makes the use of differentiating marks more reasonable. 

\begin{figure*}
  \includegraphics[width=\textwidth]{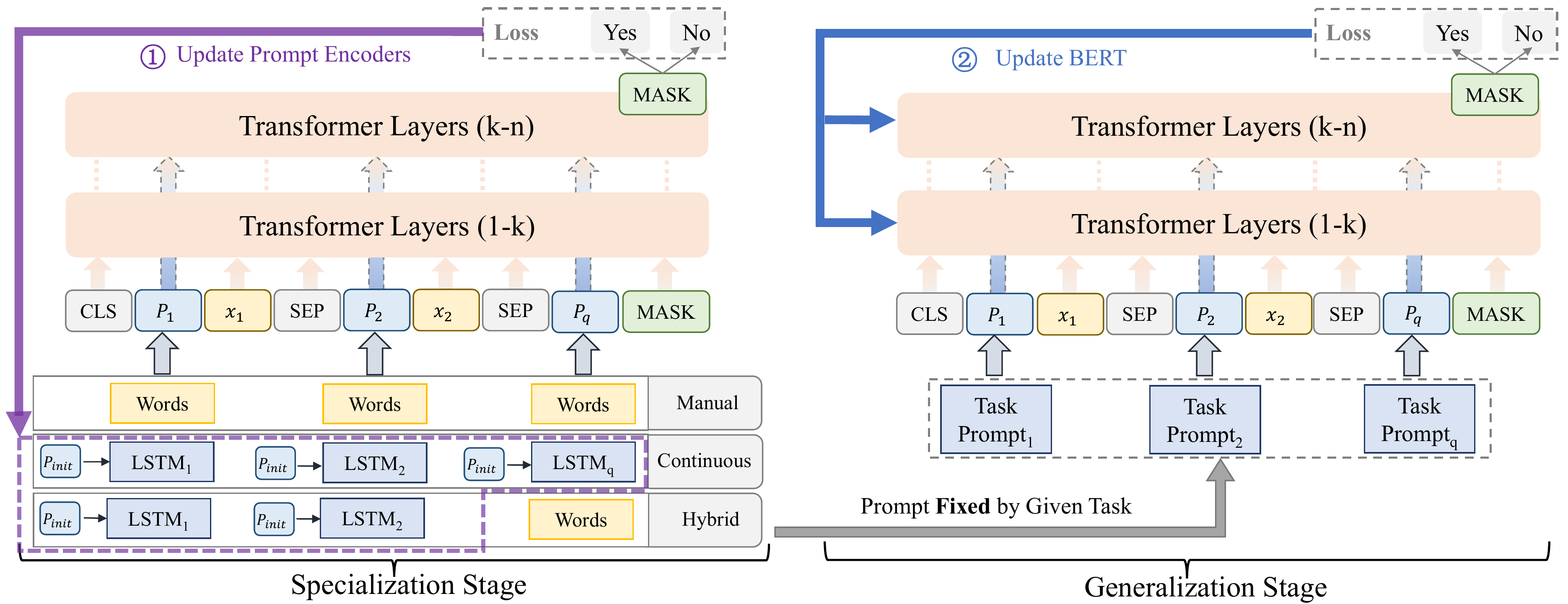}
  \caption{The design of our method. Left is the specialization stage.
  We can select manual, continuous or hybrid prompt. In manual prompt, prompts can be used directly in the generalization stage without any training. In continuous and hybrid prompt, we should train prompt encoders on each specific task and the outputs, namely $P_1$, $P_2$, $P_q$, are the prompts of the corresponding task ($P_q$ is the natural language in hybrid prompt) and are fixed in Transformer Layers (1-k). Right is generalization stage. The prompts of each task in the specialization stage are used as the prompts of the corresponding tasks in this stage.}
  \label{figure:method}
\end{figure*}

\subsection{Specialization Stage: Prompt Engineering}
In prompt engineering, we design three methods to map the description of the task to prompt tokens including manual prompt, continuous prompt, and hybrid prompt. To facilitate the description of these methods, we give a formal description of the input texts. For a pair of input texts $(x_1,x_2) $, the template of text matching is:
\begin{equation}
x_{input} = [CLS]\,P_1\,x_1\,[SEP]\,P_2\,x_2\,[SEP]\,P_q\,[MASK].\label{x_input}
 \end{equation}
The prompt tokens corresponding to each task are $P_1$, $P_2$ and $P_q$ in $x_{input}$, which can help the model distinguish different tasks in the mixed training stage. They control PLM to use the knowledge in pre-training to predict the word at [MASK] and complete the corresponding text matching task, which can better reflect the key factors of different tasks in the calculation process of PLM, so as to provide the strong differentiating effect. 

\subsubsection{\textbf{Manual Prompt}}
Inspired by LAMA~\cite{lama} that provides manually created cloze templates to probe knowledge in PLMs, manual prompt is
proposed to perform many NLP tasks such as text classification~\cite{pet} and text generation~\cite{ge-pet}. The manual prompts designed for different text matching tasks are shown in Table~\ref{table:manual-prompt}. In manual prompt, prompt tokens are entirely composed of natural language. From the left to right, $P_1$ and $P_2$ are used to tell BERT what type of $x_1$ and $x_2$ are respectively. $P_q$ is an interrogative sentence that asks questions about the relationship between 
two texts according to the specific task. And the answers to these questions can only be “yes” or “no”, which are expected to be filled in [MASK]. In addition to texts, punctuation marks such as colon and question mark also play an important role in the description of the task. Compared with continuous and hybrid prompt, this method uses natural language as prompts without additional training on prompt tokens but requires human experience and the result is sub-optimal.

\begin{table*}[t]
  \caption{Design of manual prompts. Handcrafted prompts are described in natural language.}
  \label{table:manual-prompt}
  \begin{tabular}{llll}
    \toprule
    Task & $P_1$ & $P_2$ & $P_q$\\
    \midrule
   Document Retrieval & Query: & Passage: & Dose the passage include the content matches the query?\\
    Question and Answer & Question: & Passage: & Does the passage include the answer of the question?\\
    Retrieval-based Dialogue & The first sentence: & The second sentence: & Can the second sentence reply to the first sentence?\\
    Paraphrase Identification & The first sentence: & The second sentence: & Do these two sentences mean the same thing?\\
    Natural Language Inference & Premise: & Hypothesis: & Can the hypothesis be concluded from the premise?\\
  \bottomrule
\end{tabular}
\end{table*}

\subsubsection{\textbf{Continuous Prompt}} \label{cp}

Based on P-tuning~\cite{p-tunning}, we propose a method to automatically search prompts in continuous space. In this method, prompt is a trainable continuous vector. Different from P-tuning, our method focuses on improving the multi-task generalization capability of model in text matching. Our purpose of obtaining the prompt tokens of each task is not to improve its performance in prompt learning but to let the prompt tokens give better descriptions of each specific text matching task and use these descriptions to distinguish different tasks. Therefore, different from P-tuning, we make the following improvements:

\textbf{Prompt encoder: }After pre-training, the word embedding in BERT has been very discrete. It means that if we just optimize the word embedding of the prompt tokens in BERT, the search space will be very small, and it is easy to fall into the local optimal point. So, a bidirectional long-short term memory network (LSTM) is used to optimize all prompt token embedding in a completely continuous space~\cite{p-tunning}. Different from this, we use LSTM for $P_1$, $P  _2$, and $P_q$ respectively. It is because LSTM assumes that there is a certain dependency between various tokens of prompt. That is, in bidirectional LSTM, the tokens of $P_1$, $P_2$, and $P_q$ are interdependent. But this unnecessary dependency limits the search space of vectors. So using separate LSTM for $P_1$, $P  _2$, and $P_q$ can reduce the dependency between these three prompts and expand the search space.

\textbf{Fix the prompt tokens: }In P-tuning and many other prompt learning methods, prompt tokens are just added to input text and alterable like other tokens in self-attention. Different from this, we let the embedding of trainable prompt tokens remain fixed in layers $1$ to $k$ in the self-attention of these layers, prompt tokens can affect the calculation of BERT on input text, but the input text cannot affect prompt tokens. In the layers $k$ to $n$, they are alterable. It is because if we do not impose any restrictions on prompt tokens, the embedding of the prompt tokens is affected by other input words and updates in every layer. This is not conducive to obtaining an abstract description of specific text matching task, because there is no guarantee that the embeddings of prompt tokens are only determined by the prompt encoder due to the influence of other input words, which may cause the embedding to fit the data instead of the task. While on the other hand, it is inconsistent with the mask prediction task in the pre-training process of BERT if we fix prompt tokens in each layer, which may lead to sub-optimal results when using language model to predict the word at [MASK]. Therefore, an intermediate layer needs to be determined to balance the two to achieve optimal performance. Based on our experiment, we find that the $11$-$th$ layer is an optimal boundary layer. We balance the description of the task and the interaction with other words in the text. In this way, prompt tokens can achieve better control of BERT and describe the task more comprehensively. The parameters of prompt encoders are updated through back-propagation thereby adjusting the embedding of the prompt tokens.

Let $P_{init}$ is a randomly initialized vector used as the input to prompt encoder. $MLP_1(LSTM_1)$, $MLP_2(LSTM_2)$, $MLP_q(LSTM_q)$ are the encoders of the corresponding prompt. Each encoder consists of a bidirectional LSTM and multilayer perceptron. These three encoders are optimized to obtain the continuous prompts respectively. Take $P_1$ as an example, the embedding is obtained by:
\begin{equation}
P_1 = MLP_1(LSTM_1(P_{init})).
\end{equation}
Given the $x_{input}$ and label $y$, the loss function can be expressed as $Loss(\mathcal{M}(x_{input},y))$ where $\mathcal{M}$ is the per-trained language model. $\theta^*$ are the parameters of prompt encoder, the optimization of $\theta^*$ is:
\begin{equation}
\theta^* = \mathop {argmin}_{\theta}Loss(\mathcal{M}(x_{input},y)).
\end{equation}
It is worth noting that in the training of prompt encoder, the parameters of BERT are fixed. The architecture of continuous prompt is shown in the left of Figure~\ref{figure:method}. 

\subsubsection{\textbf{Hybrid Prompt}} \label{hp}

In hybrid prompt, for $P_q$, instead of using prompt encoder to generate a continuous vector, we express it as a natural language: "Do these two sentences match?". It is because in the template of continuous prompt, text matching is converted into predicting the word at [MASK]. If the result is “yes”, these two texts match, otherwise they do not. In order to predict the word, the parameters of the three prompt encoders for $P_1$, $P_2$, and $P_q$ need to be updated. However, there is no prior information about the task other than training data. This leads to that although the prompt encoder can minimize loss during training, the task it describes may not be text matching. It just makes prompts control PLMs to extract features from the input texts, and output the corresponding "yes" or "no", which is not conducive to generalizing to other datasets. Expressing $P_q$ as task-relevant natural language can control the task closer to text matching and facilitate $P_1$ and $P_2$ to describe different matching tasks, which is beneficial for improving generalization ability. The template contains both manual prompt ($P_q$) and continuous prompt ($P_1, P_2$). We call this method hybrid prompt. The training architecture is shown in the left of Figure~\ref{figure:method}. 

\subsection{Generalization Stage: Mixed Training}

After prompt engineering in specialization stage, the descriptions of each specific task are mapped to prompt tokens. The second stage is generalization, in this stage, we guide BERT to learn the common and essential signals of text matching through the mixed datasets of multiple tasks, so as to improve its multi-task generalization ability. In addition, prompt tokens help multi-task models distinguish different tasks, adapt to task-specific matching signals and achieve good generalization. 

We mix the datasets from the different tasks and tune BERT on these mixed datasets.
The prompt tokens of each task come from specialization stage, which can be manual, continuous or hybrid prompt. We add the prompt tokens to the input text of the corresponding task according to the template shown in $x_{input}$. In order to ensure that the multi-task model is not biased towards a certain task, we sample the dataset of each task to achieve the relative balance of the amount of data for each task. Besides, we rearrange the data for each task, so that each epoch can see roughly the same number of samples from each task. We evaluate the performance every epoch and use early stopping to stop training when the performance is lower than the best result in history for 5 consecutive epochs. In order to make the roles of $P_1$, $P_2$, and $P_q$ in each layer of BERT consistent with the first stage, their calculations should be the same as the corresponding three prompt engineering methods. In this stage, all parameters of BERT are trainable. When using the model for predicting, we only need to follow this paradigm and add different prompts according to different tasks to achieve multi-task text matching by single model without any task-specific layers. The architecture is shown in the right of Figure~\ref{figure:method}.

\subsection{Training and Evaluation} \label{specialization} 
We use prompt learning to train Match-Prompt, the key of which is updating the parameters to maximize the probability that the language model correctly predicts the word at [MASK]. Let $\mathcal{M}$ be the pre-trained language model. The vocabulary of $\mathcal{M}$ is $\mathbb{V}$. $\mathbb{Y}$ is the label set in this task and $y$ is each label. We can design a verbalizer to map label to the word in vocabulary and mark it $v$. If the [MASK] is filled by “yes”, the label is $1$, if it is filled by “no”, the label is $0$. $\mathcal{M}(w|x_{input})$ is used to represent the probability of filling [MASK] by $w\in\mathbb{V}$. Relationship between $x_{input}$ and $y$ is modeled as:
\begin{equation}
s(y|x_{input}) = \mathcal{M}(v(y)|x_{input}),
\end{equation}
\begin{equation}
P(y|x_{input}) = \frac{e^{s(y|x_{input})}}{\sum\nolimits_{y' \in \mathbb{Y}}e^{s(y'|x_{input})}}.
\end{equation}
The loss function of our method is cross entropy:
\begin{equation}
\mathcal{L} = -\sum_{i=1}^{N}[y_{i}\log P_{i}+(1-y_{i}) \log (1-P_{i})].
\end{equation}
In the continuous and hybrid prompt of specialization stage, the trainable component is prompt encoder and PLM is fixed. In generalization stage, the trainable component is PLM and the prompt tokens of each task come from specialization stage. We can directly combine these prompt tokens and texts into $x_{input}$ and input them to PLM for training and predicting without any prompt encoders. 

After training, we can get the matching function $F$, the matching process can be simplified to:
\begin{equation}
y = F(x_1,x_2),
\end{equation}
$y \in \{0,1\}$, indicates whether $x_1$ and $x_2$ match. In ranking tasks such as DR, QA and RD, we use the probability that the word of [MASK] is predicted to be “yes” minus the probability that it is predicted to be “no” as the relevance between texts:
\begin{equation}
REL = \mathcal{M}(yes|x_{input})-\mathcal{M}(no|x_{input}).
\end{equation}

\section{Experiments} \label{exp}
In this section, we introduce our experiment settings and results.
\subsection{Experiment Settings}
\subsubsection{\textbf{Research Questions}}
\begin{figure}
  \centering
  \includegraphics[width=\linewidth]{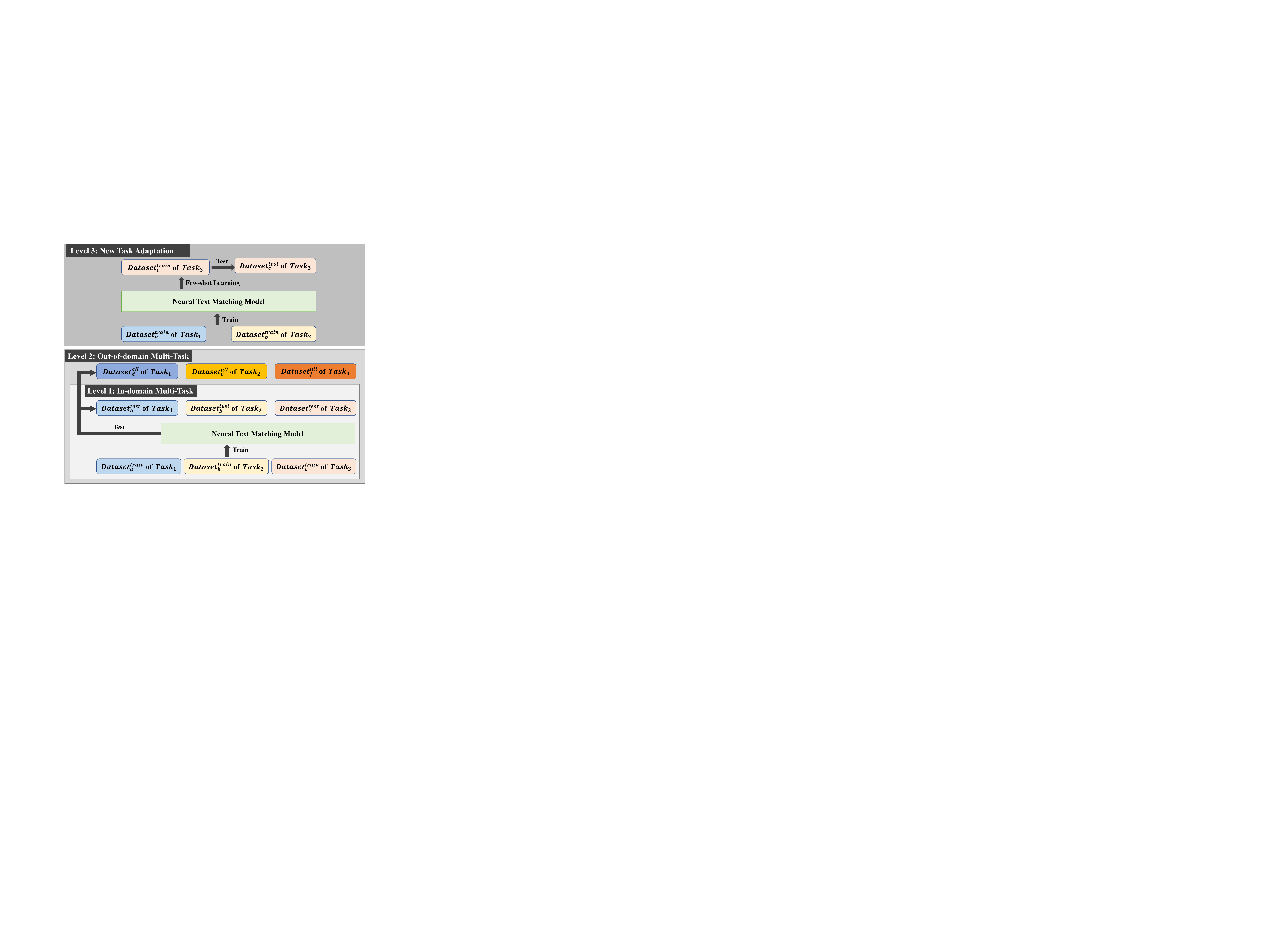}
  \caption{Three levels of multi-task generalization.}
  \label{figure:3-level}
\end{figure}

In Section~\ref{introduction}, multi-task generalization is divided into three levels. We compare the performance of Match-Prompt with the traditional fine-tuning paradigm at these three levels in experiment. Figure~\ref{figure:3-level} describes these three levels.
\begin{itemize}
\item \textbf{In-domain Multi-task Performance.} 
In order to test the ability of multi-task model to distinguish
and optimize different tasks and utilize the shared information of
tasks, given mixed datasets of multiple tasks, the model performs multi-task learning on the training set and we test its performance on each task on the testing set. 
\item \textbf{Out-of-domain Multi-task Performance. } 
In order to investigate the out-of-domain generalization ability of multi-task models, we test the zero-shot learning ability of the multi-task model on unseen datasets.
\item \textbf{New Task Adaptation. } 
In order to investigate the ability of multi-task model to use information from other tasks to facilitate low-resource adaptation, we use leave-one-out to perform few-shot learning on new tasks. 
\end{itemize}
\subsubsection{\textbf{Datasets}} The mixed datasets consist of six datasets from five tasks including Document Retrieval, Open-domain Question Answering, Retrieval-based Dialogue, Paraphrase Identification, and Natural Language Inference. We have sampled the datasets of each task to maintain a relative balance between each task. The details of mixed datasets are listed in Table~\ref{table:basic-datasets}. After getting the model, we observe its performance on the test dataset of each task. Besides, in order to explore the out-of-domain generalization capability of multi-task text matching model, we test the performance on the datasets that the model has not been trained on. These datasets are listed in Table~\ref{table:cross-datasets}. To ensure the same amount of data as the baseline, we keep the data used for training continuous and hybrid prompt in prompt engineering consistent with the generalization stage. 

\subsubsection{\textbf{Training}}
The model used in the experiment is BERT-base (109M)\footnote{https://huggingface.co/bert-base-uncased/tree/main}. The gradient descent method Adam~\cite{adam} with learning rate $10^{-5}$ is used to optimize the objective and batch size is 16. We evaluate the performance of every epoch and use early stopping to avoid overfitting. In continuous and hybrid prompt, the lengths of $P_1$, $P_2$ and $P_q$ (only in continuous prompt) are 6, 6, 5 respectively.

\begin{table}
\setlength\tabcolsep{1.5pt}%调列距
  \caption{Details of each task in mixed training datasets.}
  \label{table:basic-datasets}
  \scalebox{0.9}{
  \begin{tabular}{llll}
    \toprule
    Task & Dataset & Train(sampled) & Test\\
    \midrule
    DR & MQ2007 (MQ07)~\cite{mq2008}& 40,000 (q-d pairs) & 338 (queries)\\
    QA & TrivaQA (Triva)~\cite{trivia}& 40,000 (q-a pairs) & 8,837 (questions)\\
    RD & DailyDialog (DG)~\cite{daily}& 40,000 (dialogue pairs)& 3,209 (dialogues)\\
    PI & QQP\tablefootnote{https://www.kaggle.com/c/quora-question-pairs}+MSRP~\cite{mrpc}& 35,924+4,076 & 50,016+1,752\\
    NLI& MultiNLI (MNLI)~\cite{mnli}& 40,000 & 39,296\\
  \bottomrule
\end{tabular}
}
\end{table}

\begin{table}
  \caption{Datasets to test the out-of-domain multi-task ability.}
  \label{table:cross-datasets}
  \centering
  \scalebox{0.9}{
  \begin{tabular}{ll}
    \toprule
    Task & Datasets \\
    \midrule
    DR & \makecell[l]{ClueWeb09-B (CW)~\cite{clueweb}, Robust04 (RB04)~\cite{robust04}, Gov2\tablefootnote{https://ir-datasets.com/gov2.html}} \\
    NLI & SNLI~\cite{snli}, SICK-E~\cite{sick} \\
    QA & \makecell[l]{CuratedTREC (Trec)~\cite{trec}, Natural Questions (NQ)~\cite{nq}, \\ WebQuestions (WQ)~\cite{webq}} \\
    PI & PARADE~\cite{parade}, TwitterURL (TURL)~\cite{twitter} \\
    RD & Reddit, AmazonQA (AQ)~\cite{amazonqa}\\
    \bottomrule
\end{tabular}
}
\end{table}

\subsubsection{\textbf{Specical Pre-processing}}

For QA, we construct q-a pairs in training set as described in DPR~\cite{dpr}. In order to balance the number of positive and negative samples, we sample three positive and three negative samples for each question in training. These three negative samples consist of two hard negative and one gold passage (positive passage paired with other questions). For PI, the size of public datasets is small, so we select two datasets including MSRP and QQP. For NLI, we convert this task into a binary classification task that whether hypotheses can be inferred from the premises. For DR and QA, we use BM25 to retrieve items and use our model to re-rank them. For RD, the candidate set of the sentence consists of one positive and nine negative samples.

\subsubsection{\textbf{Baselines and Measures}}
PLMs improve the performance of interaction-based text matching. For example, MonoBert~\cite{monobert} takes the concatenating query and document as the input and feeds the embedding of [CLS] into a feed-forward network to judge the relevance, which is the traditional fine-tuning paradigm of BERT. In experiment, we want to show that the text matching model trained with Match-Prompt can get better multi-task generalization ability than the traditional fine-tuning paradigm. Some methods~\cite{monobert,plminir,qa_bert_sota,pi-bert,MT-DNN,match-ignition,cedr} are not considered because they are just variants based on this fine-tuning paradigm for different matching tasks, which are inconsistent with our motivation. So the most important baseline in the experiment is Fine-tuning.

As for the multi-task learning method for Fine-tuning, we introduce three methods. \textbf{Fine-tuning$_{multi}$}, which uses traditional fine-tuning paradigm to train BERT without any task-specific marks. \textbf{Fine-tuning$_{mark}$}, which adds the task-specific marks to the input text.
\textbf{MT-DNN}~\cite{MT-DNN} adds the task-specific feed-forward networks for each task, we reproduce it on our mixed datasets. This method introduces additional parameters, and the number of parameters increases with the number of tasks, but Match-Prompt does not need any task-specific layers during inference. There is also a multi-task learning framework that converts each task into a unified question answering format~\cite{zero-shot-tc,tc,MQAN,t5}. We reproduce this framework on our mixed datasets using 
GPT2-base (117M)\footnote{https://huggingface.co/gpt2} and T5-base (220M)\footnote{https://huggingface.co/google/t5-v1\_1-base} and call them \textbf{MTL$_{GPT2}$} and \textbf{MTL$_{T5}$} respectively. Since T5 has been pre-trained on multiple supervised downstream tasks that will be tested in our experiment, which is unfair for comparison, we choose T5 1.1 that is only pre-trained on unsupervised datasets. We also reproduce \textbf{MQAN}~\cite{MQAN} on our tasks, which is a classic QA-based multi-task model. \textbf{BM25}~\cite{bm25} also has strong multi-task generalization ability, we use it as one of the baselines. In multi-task training, there are some tricks about data sampling, loss construction and task scheduling, which can be used by both baselines and Match-Prompt, so we do not compare them in detail. In order to compare the performance of the multi-task model enhanced by Match-Prompt with the task-specific model, the models (i.e. \textbf{Fine-tuning$_{sp}$}) are specifically trained on the dataset corresponding to each task listed in Table~\ref{table:basic-datasets}, and tested on the corresponding task.

For our method, \textbf{Match-Prompt$_{mp}$, Match-Prompt$_{cp}$, Match-Prompt$_{hp}$} correspond to manual, continuous and hybrid
prompt in text matching described in Section~\ref{methods}.

We use Accuracy and F1-score to evaluate NLI and PI~\cite{match,esim}. For QA and RD, we use P@1 and MRR~\cite{SRNN}. For DR, we use NDCG~\cite{prop}.

\subsection{Main Results}

\subsubsection{\textbf{In-domain Multi-task Performance}} \label{basic-exp}
We train multi-task model on the mixed datasets listed in Table~\ref{table:basic-datasets} and test the performance of the model on each specific task. The results are shown in Table~\ref{table:basic-results}. Compared with other multi-task models, Match-Prompt shows improvement in multi-task text matching. All three prompt engineering methods exceed other multi-task models in most tasks, and hybrid prompt achieves the best results on most datasets. It is worth noting that Match-Prompt$_{hp}$ surpasses Fine-tuning$_{sp}$ in every task. We can infer from these results that Match-Prompt can improve the in-domain multi-task performance of BERT in text matching. Match-Prompt enables BERT to better utilize the matching signals shared by each task and the differentiating marks of each task are added to the input text in the form of prompts to predict the word at [MASK], which makes the use of differentiating marks more reasonable, thereby improving its multi-task performance.

\begin{table*}[htbp]
  \caption{Performance of multi-task model trained on the mixed datasets listed in Table~\ref{table:basic-datasets}. Boldface indicates the best results of multi-task models and the results over Fine-tuning$_{sp}$ are denoted as `$*$'. Results with significant performance improvement with p-value $ \leq 0.05$ compared with all multi-task models of baselines are denoted as `$+$'.}
  \label{table:basic-results}
\renewcommand\arraystretch{1}
%% 设置表格每一行
\setlength\tabcolsep{4pt}%调列距
\renewcommand{\cellset}{\renewcommand{\arraystretch}{0.5}} 
%% 设置单元格内行间距
\centering
   \scalebox{0.9}{
\begin{tabular}{llllllllll}
\toprule
\multirow{2}{*}{Method} & \multicolumn{2}{c}{Trivia$_{QA}$} & \multicolumn{2}{c}{DailyDialog$_{RD}$} & \multicolumn{2}{c}{MSRP+QQP$_{PI}$}  & \multicolumn{2}{c}{MultiNLI$_{NLI}$} & MQ2007$_{DR}$ \\ 
                        &  P@1(\%)           & MRR(\%)             & P@1(\%)           & MRR(\%)               & Acc.(\%)           & F$_1$(\%)             & Acc.(\%)           & F$_1$(\%)            & NDCG@10(\%) \\ \hline  \hline 
\multicolumn{10}{c}{Task-specific model for each specific task} \\    
Fine-tuning$_{sp}$  & 47.17          & 55.62          & 78.59            & 84.86            & 83.42          & 79.66  & 84.40            & 77.35          & 50.60   \\  \hline  
\multicolumn{10}{c}{Multi-task model for all tasks} \\
BM25   & 46.88          & 54.67          & 35.81             & 50.39           & 70.31          & 77.62 & 66.37           & 57.23          & 41.40        \\
MQAN      & \makecell[l]{7.33}          & \makecell[l]{16.72}          & \makecell[l]{68.06}             & \makecell[l]{74.92}            & \makecell[l]{60.00}          & \makecell[l]{46.64}         & \makecell[l]{61.72}            & \makecell[l]{50.81}           & \makecell[l]{30.35} \\

MTL$_{GPT2}$      & \makecell[l]{34.02}          & \makecell[l]{43.29}          & \makecell[l]{58.58}             & \makecell[l]{70.12}            & \makecell[l]{65.20}          & \makecell[l]{62.15}         & \makecell[l]{67.59}            & \makecell[l]{58.37}           & \makecell[l]{45.72} \\

MTL$_{T5}$      & \makecell[l]{36.24}          & \makecell[l]{46.63}          & \makecell[l]{62.49}             & \makecell[l]{74.99}            & \makecell[l]{71.54}          & \makecell[l]{74.70}         & \makecell[l]{77.28}            & \makecell[l]{64.73}           & \makecell[l]{48.63} \\

MT-DNN      & \makecell[l]{46.01}          & \makecell[l]{54.46}          & \makecell[l]{75.10}             & \makecell[l]{82.05}            & \makecell[l]{81.50}          & \makecell[l]{77.10}         & \makecell[l]{82.33}            & \makecell[l]{73.92}           & \makecell[l]{47.94} \\ 

Fine-tuning$_{multi}$      & \makecell[l]{46.19}          & \makecell[l]{54.95}          & \makecell[l]{74.29}             & \makecell[l]{81.47}            & \makecell[l]{80.85}          & \makecell[l]{77.13}         & \makecell[l]{81.45}            & \makecell[l]{72.77}           & \makecell[l]{47.59} \\

Fine-tuning$_{mark}$      & \makecell[l]{46.05}          & \makecell[l]{54.72}          & \makecell[l]{74.80}             & \makecell[l]{81.95}            & \makecell[l]{81.47}          & \makecell[l]{75.44}         & \makecell[l]{82.55}            & \makecell[l]{74.01}           & \makecell[l]{49.94} \\
\hdashline

Match-Prompt$_{mp}$     & \makecell[l]{47.67$^{*}_{+}$}          & \makecell[l]{55.87$^{*}_{+}$}          & \makecell[l]{78.00$_{+}$}             & \makecell[l]{84.47$_{+}$}            & \makecell[l]{82.78$_{+}$} & \makecell[l]{79.48$_{+}$}          & \makecell[l]{83.56$_{+}$}           & \makecell[l]{76.07$_{+}$}         & 51.42$^{*}_{+}$  \\

Match-Prompt$_{cp}$     & \makecell[l]{47.79$^{*}_{+}$}          & \makecell[l]{55.98$^{*}_{+}$}          & \makecell[l]{78.01$_{+}$}             & \makecell[l]{84.58$_{+}$}            & \makecell[l]{83.36$_{+}$}  & \makecell[l]{79.42$_{+}$}          & \makecell[l]{84.56$^{*}_{+}$ }           & \makecell[l]{76.82$_{+}$}    & 51.52$^{*}_{+}$      \\

Match-Prompt$_{hp}$    & \makecell[l]{\textbf{47.92$^{*}_{+}$}}  & \makecell[l]{\textbf{56.14$^{*}_{+}$}}  & \makecell[l]{\textbf{78.65$^{*}_{+}$}}    & \makecell[l]{\textbf{84.87$^{*}_{+}$}}    & \makecell[l]{\textbf{83.55$^{*}_{+}$}}           & \makecell[l]{\textbf{79.82$^{*}_{+}$}}          & \makecell[l]{\textbf{84.64$^{*}_{+}$}}    & \makecell[l]{\textbf{77.40$^{*}_{+}$}}  & \makecell[l]{\textbf{52.01$^{*}_{+}$}} \\
\toprule
\end{tabular}}
\end{table*}

\subsubsection{\textbf{Out-of-domain Multi-task Performance}}
\begin{table*}[htbp]
  \caption{Multi-task generalization ability of the model on unseen datasets listed in Table~\ref{table:cross-datasets}. Boldface indicates the best results of multi-task models and the results over Fine-tuning$_{sp}$ are denoted as `$*$'. Results with significant performance improvement with p-value $ \leq 0.05$ compared with all multi-task models of baselines are denoted as `$+$'. }
  \label{table:cross-datasets-results}
\renewcommand\arraystretch{1.2}
%% 设置表格每一行行距
\setlength\tabcolsep{1.5pt}%调列距
\renewcommand{\cellset}{\renewcommand{\arraystretch}{0.5}} 
%% 设置单元格内行间距
\centering
   \scalebox{0.75}{
\begin{tabular}{llllllllllllllllllllllll}
\toprule
\multirow{2}{*}{Method} & \multicolumn{2}{c}{SNLI$_{NLI}$} &\multicolumn{2}{c}{SICK-E$_{NLI}$} & \multicolumn{2}{c}{PARADE$_{PI}$}   & \multicolumn{2}{c}{TURL$_{PI}$} & \multicolumn{2}{c}{Reddit$_{RD}$} & \multicolumn{2}{c}{AQ$_{RD}$} & \multicolumn{2}{c}{Trec$_{QA}$} & \multicolumn{2}{c}{NQ$_{QA}$} & \multicolumn{2}{c}{WQ$_{QA}$} & CW$_{DR}$      & RB04$_{DR}$  & Gov2$_{DR}$ \\ 
                        & Acc.           & F$_1$             & Acc.           & F$_1$           & Acc.           & F$_1$  & Acc.           & F$_1$ & P@1           & MRR  & P@1           & MRR    & P@1           & MRR             & P@1           & MRR  & P@1           & MRR  & \multicolumn{3}{c}{NDCG@10}\\ \hline \hline 
\multicolumn{21}{c}{Task-specific model for each specific task} \\
Fine-tuning$_{sp}$            & 82.19          & 74.21          & 81.61            & 73.12           & 68.34         & 60.85   &83.33 &60.45 &49.64
 &64.13 & 77.64 & 85.22 & 47.95          & 62.19          & 22.97          & 36.24          & 36.52             & 51.50           & 26.01         & 40.28 & 43.27 \\  \hline
\multicolumn{21}{c}{Multi-task model for all tasks } \\   
BM25             & 66.47    & 58.45      & 78.22    & 69.31    & 62.07  & 54.37   & 81.68  & 59.72  & 24.63 & 44.81 & 69.73 & 76.08 & 31.94          & 46.18      & 24.75$^{*}$    & 37.30$^{*}$ &  26.47          & 39.17            & 23.00         & 40.20 & 41.70\\ 
MQAN             & \makecell[l]{61.02}          & \makecell[l]{40.05}          & \makecell[l]{66.32}             &\makecell[l]{47.42}          & \makecell[l]{55.08}         & \makecell[l]{44.75}    &\makecell[l]{68.82} &\makecell[l]{55.49} &\makecell[l]{25.20}  &\makecell[l]{44.71}   
&\makecell[l]{19.80}  &\makecell[l]{32.25}  & \makecell[l]{11.57}          & \makecell[l]{25.61}           & \makecell[l]{2.24}          & \makecell[l]{9.45}          & \makecell[l]{9.54}             & \makecell[l]{20.63}           & \makecell[l]{11.98}                   & \makecell[l]{14.89} & \makecell[l]{22.54} \\

MTL$_{GPT2}$             & \makecell[l]{69.17}          & \makecell[l]{34.46}          & \makecell[l]{66.71}             &\makecell[l]{51.89}          & \makecell[l]{66.71}         & \makecell[l]{57.49}    &\makecell[l]{80.72} &\makecell[l]{59.50} &\makecell[l]{30.85}  &\makecell[l]{47.52}   
&\makecell[l]{68.79}  &\makecell[l]{75.67}  & \makecell[l]{31.75}          & \makecell[l]{47.23}           & \makecell[l]{13.17}          & \makecell[l]{25.42}          & \makecell[l]{19.63}             & \makecell[l]{34.12}           & \makecell[l]{23.27}                   & \makecell[l]{37.62} & \makecell[l]{45.12$^{*}$} \\

MTL$_{T5}$             & \makecell[l]{69.56}          & \makecell[l]{34.72}          & \makecell[l]{67.60}             &\makecell[l]{52.51}          & \makecell[l]{69.81$^{*}$}         & \makecell[l]{59.85}    &\makecell[l]{83.83$^{*}$} &\makecell[l]{62.53$^{*}$}  &\makecell[l]{33.96}  &\makecell[l]{50.35}   
&\makecell[l]{72.49}  &\makecell[l]{79.95}  & \makecell[l]{34.88}          & \makecell[l]{50.34}           & \makecell[l]{14.42}          & \makecell[l]{26.61}          & \makecell[l]{20.70}             & \makecell[l]{36.09}           & \makecell[l]{25.75}                   & \makecell[l]{39.90} & \makecell[l]{46.89$^{*}$} \\

MT-DNN         & \makecell[l]{77.85}          & \makecell[l]{70.30}          & \makecell[l]{79.11}             & \makecell[l]{70.54}           & \makecell[l]{65.10 }         & \makecell[l]{49.68 }  & 82.55 & 52.17 & 55.72$^{*}$ & 68.36$^{*}$ & 75.98 & 83.47  & \makecell[l]{41.55 }          & \makecell[l]{57.09 }          & \makecell[l]{21.27 }          & \makecell[l]{34.56}          & \makecell[l]{32.34}             & \makecell[l]{48.52}   & \makecell[l]{25.16} & \makecell[l]{40.18} & \makecell[l]{47.09$^{*}$}       \\

Fine-tuning$_{multi}$             & \makecell[l]{78.24}          & \makecell[l]{66.73}          & \makecell[l]{77.57}             &\makecell[l]{68.79}          & \makecell[l]{64.95}         & \makecell[l]{60.27}    &\makecell[l]{76.25} &\makecell[l]{50.47} &\makecell[l]{50.95$^{*}$}  &\makecell[l]{66.63$^{*}$}   
&\makecell[l]{73.09}  &\makecell[l]{80.21}  & \makecell[l]{45.55}          & \makecell[l]{60.12}           & \makecell[l]{22.22}          & \makecell[l]{36.01}          & \makecell[l]{32.27}             & \makecell[l]{48.49}           & \makecell[l]{26.05$^{*}$}                   & \makecell[l]{40.89$^{*}$} & \makecell[l]{47.05$^{*}$} \\

Fine-tuning$_{mark}$             & \makecell[l]{80.69}          & \makecell[l]{{71.45}}          & \makecell[l]{79.06 }             &\makecell[l]{67.81}          & \makecell[l]{67.67}         & \makecell[l]{60.71}    &\makecell[l]{81.39} &\makecell[l]{59.06} &\makecell[l]{56.20$^{*}$}  &\makecell[l]{69.75$^{*}$}   
&\makecell[l]{76.85}  &\makecell[l]{84.66} & \makecell[l]{45.82}          & \makecell[l]{60.71}  & \makecell[l]{25.81$^{*}$}          & \makecell[l]{39.82$^{*}$}          & \makecell[l]{34.82}             & \makecell[l]{50.17}           & \makecell[l]{25.56}                   & \makecell[l]{40.21} & \makecell[l]{47.17$^{*}$}  \\
 \hdashline

Match-Prompt$_{mp}$     & \makecell[l]{82.01$_{+}$}          & \makecell[l]{72.71$_{+}$ }          & \makecell[l]{79.14}             & \makecell[l]{70.50}           & \makecell[l]{69.24$^{*}$}         & \makecell[l]{62.15$^{*}$}  & 84.82$^{*}_{+}$ & 63.43$^{*}_{+}$ & 57.30$^{*}_{+}$ & 70.95$^{*}_{+}$ & 79.40$^{*}_{+}$ & 86.24$^{*}_{+}$  & \makecell[l]{48.89$^{*}_{+}$}          & \makecell[l]{63.02$^{*}_{+}$}          & \makecell[l]{\textbf{26.21$^{*}_{+}$} }          & \makecell[l]{\textbf{40.09$^{*}_{+}$}}          & \makecell[l]{35.38$_{+}$}             & \makecell[l]{50.32$_{+}$}           & \makecell[l]{\textbf{26.93$^{*}_{+}$}}  & \makecell[l]{41.29$^{*}_{+}$} & \makecell[l]{\textbf{48.53$^{*}_{+}$}}       \\

Match-Prompt$_{cp}$     & \makecell[l]{82.06$_{+}$}          & \makecell[l]{73.91$_{+}$}          & \makecell[l]{81.53$_{+}$}              & \makecell[l]{73.04$_{+}$}             & \makecell[l]{70.01$^{*}_{+}$}         & \makecell[l]{62.20$^{*}_{+}$}    & 84.75$^{*}_{+}$ & 63.98$^{*}_{+}$ &57.70$^{*}_{+}$ &71.96$^{*}_{+}$   & 83.26$^{*}_{+}$ & 88.35$^{*}_{+}$ & \makecell[l]{48.06$^{*}_{+}$}          & \makecell[l]{62.37$^{*}_{+}$ }          & \makecell[l]{23.50$^{*}$ }           & \makecell[l]{36.81$^{*}$}          & \makecell[l]{35.90$_{+}$}              & \makecell[l]{51.31$_{+}$}              & \makecell[l]{26.41$^{*}_{+}$}  & \makecell[l]{40.54$^{*}$} & \makecell[l]{47.61$^{*}_{+}$} \\

Match-Prompt$_{hp}$     & \makecell[l]{\textbf{83.88$^{*}_{+}$}}          & \makecell[l]{\textbf{74.98$^{*}_{+}$}}          & \makecell[l]{\textbf{81.99$^{*}_{+}$}}             & \makecell[l]{\textbf{73.35$^{*}_{+}$}}           & \makecell[l]{\textbf{70.03$^{*}_{+}$} }         & \makecell[l]{\textbf{62.21$^{*}_{+}$}}  & \textbf{85.54$^{*}_{+}$} & \textbf{64.37$^{*}_{+}$} & \textbf{58.82$^{*}_{+}$} & \textbf{72.29$^{*}_{+}$} & \textbf{85.28$^{*}_{+}$} & \textbf{90.07$^{*}_{+}$}  & \makecell[l]{\textbf{49.02$^{*}_{+}$}}          & \makecell[l]{\textbf{63.13 $^{*}_{+}$}}          & \makecell[l]{25.94$^{*}_{+}$ }          & \makecell[l]{39.94$^{*}_{+}$}          & \makecell[l]{\textbf{36.64$^{*}_{+}$}}             & \makecell[l]{\textbf{51.58$^{*}_{+}$}}            & \makecell[l]{26.82$^{*}_{+}$}  & \makecell[l]{\textbf{41.44$^{*}_{+}$}} & \makecell[l]{48.21$^{*}_{+}$} \\
\toprule
\end{tabular}

}
\end{table*}
We test the multi-task model trained on the mixed datasets (listed in Table~\ref{table:basic-datasets}) on unseen datasets (listed in Table~\ref{table:cross-datasets}). The experimental results in Table~\ref{table:cross-datasets-results} indicate that Match-Prompt$_{hp}$ gets the best performance on most datasets. This experiment further demonstrates that Match-Prompt greatly improves the generalization ability of text matching models to different tasks and datasets. Prompt learning is closer to the pre-training process, so it preserves the knowledge of the BERT. In addition, prompt tokens have the description for each specific task, which facilitates migration on different datasets and training on mixed datasets consisting of multiple tasks can improve the generalization robustness of the model~\cite{multiqa}, so Match-Prompt can improve the generalization ability of the model. The performance of other multi-task models drops seriously than Fine-tuning$_{sp}$ but Match-Prompt$_{hp}$ exceeds it. This shows that our method is conducive to fusing and exploiting the knowledge of each task and avoiding interference between different tasks in multi-task learning. Match-Prompt$_{hp}$ performs better than Match-Prompt$_{cp}$, which proves that it is effective to use natural language to control continuous prompt tokens learning to describe the task.

\subsubsection{\textbf{New Task Adaptation}} \label{4.2.3}

We also explore the few-shot learning capability of the multi-task model in new tasks. For the five specific text matching tasks, we select four of them for mixed training and call this model multi-task$_{sub}$ and the remaining one is used as the new task of the multi-task$_{sub}$ (leave-one-out). In order to ensure the number of samples used by each method is consistent, in Match-Prompt$_{hp}$, the prompt tokens of the new task are obtained by the weighted sum of the prompt tokens of the other four tasks, and the weight of each task is learnable variable in few-shot learning, reflecting the similarity of the matching signals between two tasks. The w/o fusion is do not fuse the prompts of other tasks and train directly from scratch under few-shot settings. In order to better demonstrate the promotion effect of other tasks on low-resource learning, we also use Fine-tuning$_{sp}$ to directly perform few-shot learning on each task. For the new task, we select 32 positive and 32 negative samples for training and observe the performance of the model on testing set. The experimental results are shown in Table~\ref{table:few-shot-results}. The models based on multi-task$_{sub}$ get better few-shot learning performance than Fine-tuning$_{sp}$. Besides, Multi-task$_{sub}$ based on Match-Prompt has better few-shot learning ability in new tasks than that based on other methods. This indicates that the information of other tasks can improve the adaptability of the model to new tasks and Match-Prompt performs best. In addition, the prompts of other tasks contain descriptions of essential matching signals, and the integration of them can further promote few-shot learning.

\begin{table}
  \caption{ Few-shot learning performance of the models.
  Each task in this table is a new task not included in mixed datasets for multi-task$_{sub}$ (leave-one-out). Results with significant performance improvement with p-value $ \leq 0.05$ compared with all multi-task models of baselines are denoted as `$+$'.}
  \label{table:few-shot-results}
\renewcommand\arraystretch{1.1}
%% 设置表格每一行
\setlength\tabcolsep{4pt}%调列距
\renewcommand{\cellset}{\renewcommand{\arraystretch}{0.5}} 
%% 设置单元格内行间距
\centering
   \scalebox{0.9}{
\begin{tabular}{lccccl}
\toprule
\multirow{2}{*}{Method} &{Trivia} & {DG} & {MSRP}   & {MNLI} & MQ07\\
                        & MRR             & MRR          & Acc             & Acc         &NDCG@10          \\ \hline \hline
\multicolumn{6}{c}{Few-shot learning on each task} \\
Fine-tuning$_{sp}$    & \makecell[l]{24.13}  & \makecell[l]{38.97}  & \makecell[l]{64.93} & \makecell[l]{35.76} & \makecell[l]{33.87} \\       \hline
\multicolumn{6}{c}{Few-shot learning based on multi-task$_{sub}$ on each task} \\
MTL$_{GPT2}$    & \makecell[l]{31.52}          & \makecell[l]{41.26}         & \makecell[l]{69.21}             & \makecell[l]{48.53}            & \makecell[l]{34.12}\\
MTL$_{T5}$    & \makecell[l]{35.85}          & \makecell[l]{45.92}         & \makecell[l]{71.13}             & \makecell[l]{48.78}            & \makecell[l]{35.34}\\
MT-DNN   & \makecell[l]{37.39}          & \makecell[l]{50.97}          & \makecell[l]{68.17}             & \makecell[l]{59.16}            & \makecell[l]{35.12} \\ 
Fine-tuning$_{multi}$   & \makecell[l]{40.11}          & \makecell[l]{51.03}         & \makecell[l]{67.01}             & \makecell[l]{52.50}            & \makecell[l]{36.11}\\

Fine-tuning$_{mark}$   & \makecell[l]{37.49}          & \makecell[l]{50.06}          & \makecell[l]{70.90}             & \makecell[l]{59.98}            & \makecell[l]{36.02}\\

\hdashline

Match-Prompt$_{hp}$     & \makecell[l]{\textbf{45.13$_{+}$}}          & \makecell[l]{\textbf{52.59$_{+}$}}         & \makecell[l]{\textbf{74.15$_{+}$}}            & \makecell[l]{\textbf{62.02$_{+}$}}            & \makecell[l]{\textbf{37.18$_{+}$}} \\ 
- w/o fusion     & \makecell[l]{42.37$_{+}$}          & \makecell[l]{51.09$_{+}$}         & \makecell[l]{71.01$_{+}$}            & \makecell[l]{60.92$_{+}$}            & \makecell[l]{36.28$_{+}$} \\
\toprule
\end{tabular}
}
\end{table}

\subsection{Model Analysis} \label{ablation}
The quality of learned prompt tokens is very important for the multi-task generalization of text matching models. We first measure the specificity of the information stored in the prompt tokens by evaluating on handcrafted `QA vs PI' task. Then, we explore the connections between task prompt tokens and their composability.

\textbf{Ability to Distinguish Tasks.} 
We construct handcrafted `QA vs PI' datasets to verify it and the examples come from given QA datasets (Trec, WQ and NQ). For each question, we provide it with two candidate texts, one is the passage containing the answer (a positive sample in original QA datasets), denoted as $a$, and one is the question itself, denoted as $q$. Since $q$ is exactly the same as the question, while can not be the answer to the question, so it can only reflect the matching relation in PI task. In contrast, $a$ contains the answer of question and can be use to denote the matching relation in QA task. Due to the extra answer information in $a$, its matching score under PI task will be lower than $q$ which is exactly the same as question. Therefore, by altering task marks between QA and PI and comparing the matching scores of two candidates, we can evaluate the ability of multi-task model to distinguish tasks.
Table~\ref{table:dis} indicates that Match-Prompt can improve the ability of multi-task model to distinguish different tasks. Even though other multi-task learning methods distinguish tasks by different means and are trained on datasets of multiple tasks, they still tend to the exact matching signals but cannot distinguish tasks well. This is also the key reason why Match-Prompt works better.

\begin{table}
  \caption{Illustrate the ability to distinguish tasks.}
  \label{table:dis}
\renewcommand\arraystretch{1}
%% 设置表格每一行行距
\setlength\tabcolsep{2.5pt}%调列距
\renewcommand{\cellset}{\renewcommand{\arraystretch}{0.5}} 
\scalebox{0.9}{
\begin{tabular}{lllllll}
\toprule
\multirow{3}{*}{Method} & \multicolumn{3}{c}{As QA Task} & \multicolumn{3}{c}{As PI Task} \\
                        & Trec& WQ& NQ & Trec & WQ& NQ\\
                        & P$_{a>q}$     & P$_{a>q}$              & P$_{a>q}$ & P$_{q>a}$          & P$_{q>a}$   & P$_{q>a}$ \\ \hline 
MTL$_{GPT2}$             & 35.89         & 10.52         & 14.82 & 78.25        & 90.72         & 94.53 \\
MTL$_{T5}$             & 38.35         & 6.39          & 9.03 & 76.07         & 97.09          & 95.45 \\
MT-DNN             & 27.76         & 29.11          & 19.94 & 98.67         & 99.56          & 99.70 \\ 
Fine-tuning$_{multi}$             & 20.91         & 13.47          & 4.86 & 79.09         & 86.53          & 95.14 \\
Fine-tuning$_{mark}$             & 40.75         & 34.82          & 10.49 & 63.70         & 72.06          & 97.37 \\ 
\hdashline
Match-Prompt$_{hp}$  & \textbf{71.98}         & \textbf{70.68}          & \textbf{59.92}    & \textbf{100.00}         & \textbf{100.00} & \textbf{100.00}     \\  
\toprule
\end{tabular}
}
\end{table}

\begin{figure}
    \centering
    \subfigure[Cosine similarity for prompt tokens ]{
	\includegraphics[width=1.58in]{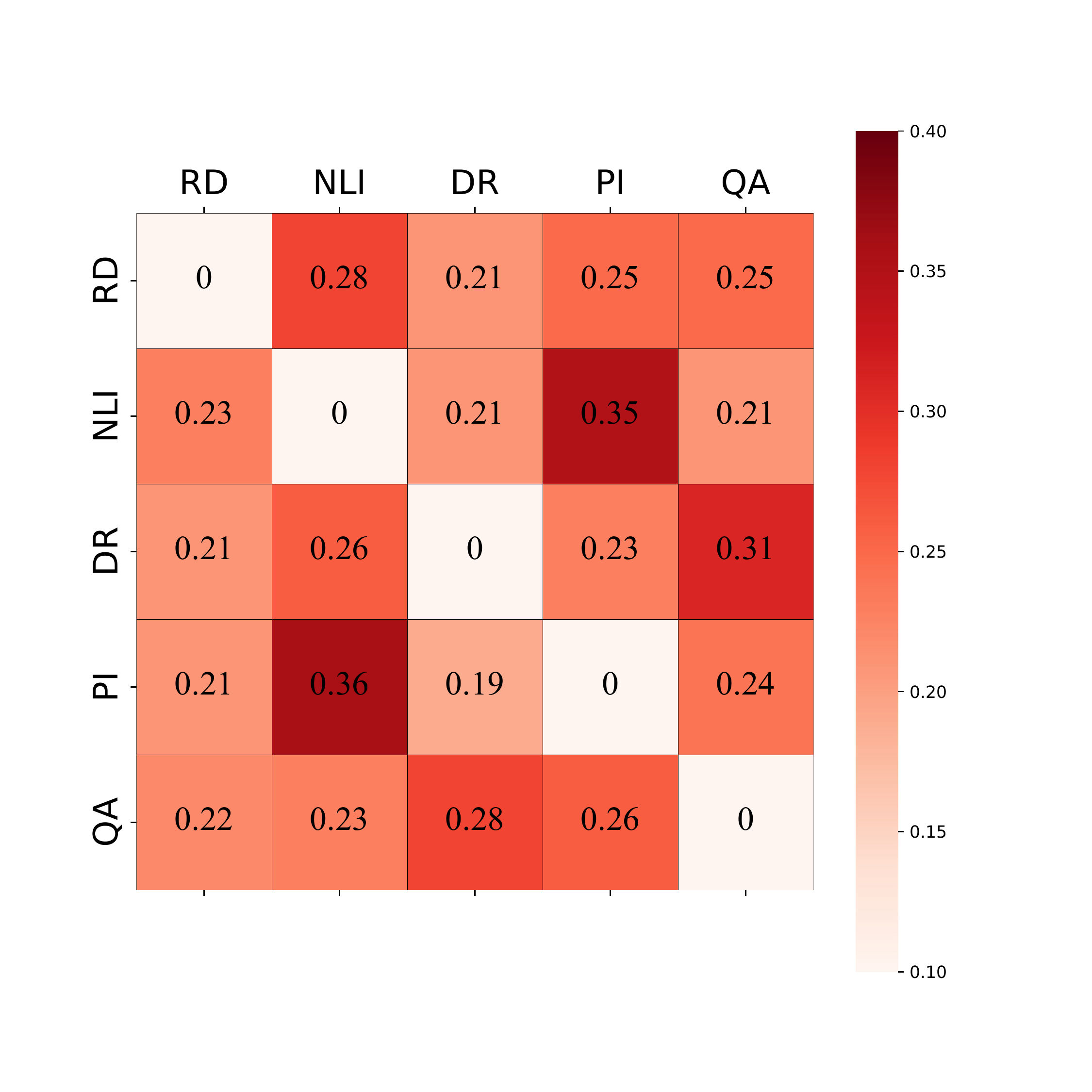}
	\label{prompt heatmap}
    }
    \subfigure[Weights for new task prompt tokens]{
        \includegraphics[width=1.58in]{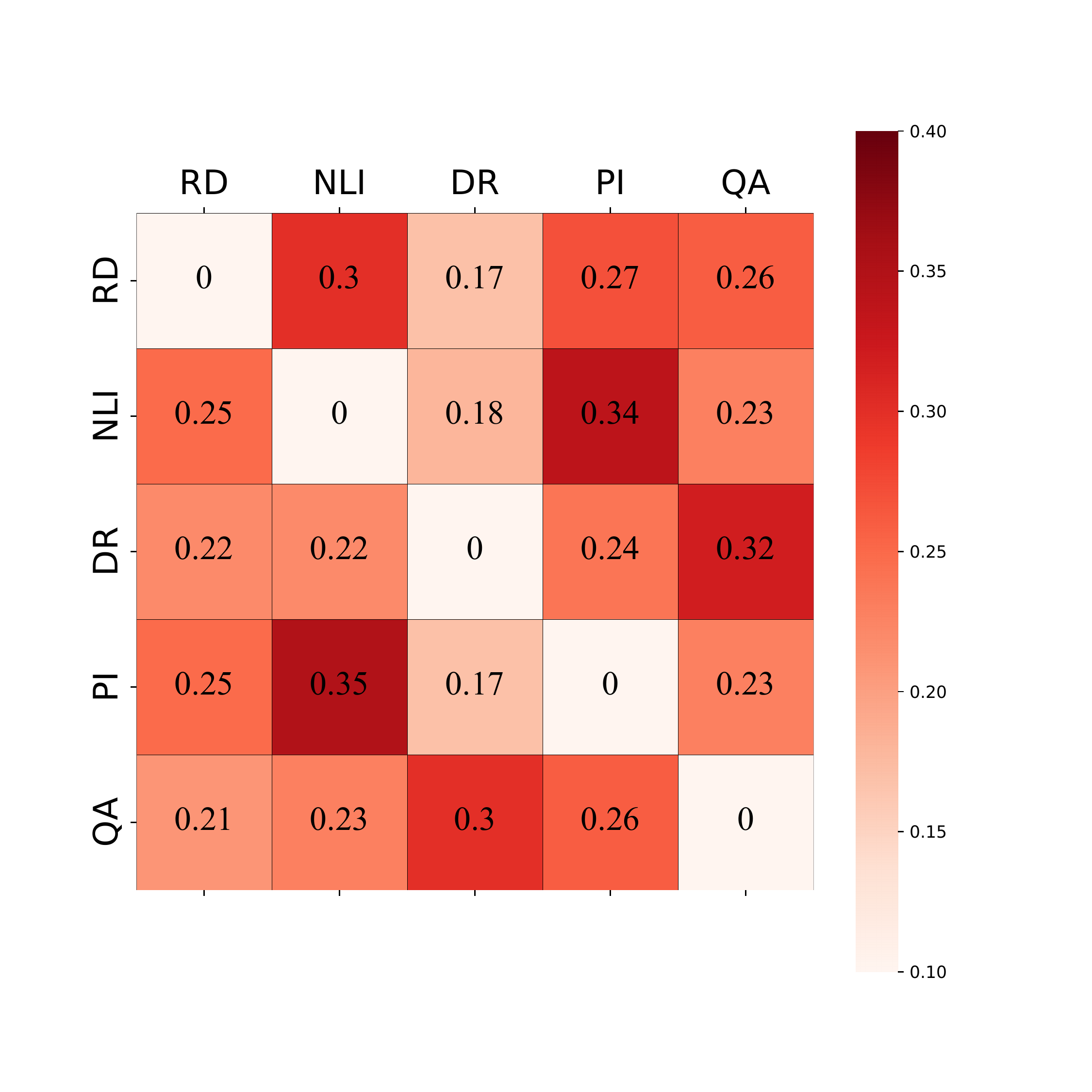}
    \label{weight}
    }
    \caption{Task relationships using prompt tokens. The sum of each row is 1, and the value (keep two decimal places) represents the proportion of the cosine similarity or weight of other tasks to the task of each row.}
    \label{heatmap}
\end{figure}

\textbf{Relationships between Tasks.}
We explore the relationships between tasks using the cosine similarity of task prompt token embeddings learned in Section~\ref{basic-exp} and its heatmap is shown in Figure~\ref{prompt heatmap}. We can see that NLI is similar to PI because they both focus on exact matching signals, DR is similar to QA because both exact and semantic matching signals are important in these tasks. The similarity between different tasks is generally consistent with our prior knowledge of the task. 
Heatmap of the fusion weights of tasks for each new task obtained in Section~\ref{4.2.3} is shown in Figure~\ref{weight}. This weight distribution is consistent with the embeddings similarity distribution of the prompt tokens between tasks, which further supports the rationality of fusing different tasks for new task adaptation.

\section{Conclusion}

We introduce prompt learning into multi-task text matching and propose a specialization-generalization training strategy, namely Match-Prompt, to improve the multi-task generalization capability of BERT in text matching tasks. In specialization stage, the descriptions of different matching tasks are mapped to only a few prompt tokens. In generalization stage, text matching model explores the essential matching signals by being trained on diverse multiple matching tasks. In the inference, the prompt tokens help the model distinguish different tasks and can be combined to adapt to new tasks based on the essential matching signals. Experimental results show that our method improves multi-task generalization ability of BERT in text matching and the performance of our multi-task model on each task can surpass the task-specific model. Our work indicates that there are not only different but also essential and common matching signals in different text matching tasks, Match-Prompt can help PLMs leverage these signals and distinguish different tasks to get stronger in-domain multi-task, out-of-domain multi-task and new task adaptation performance than traditional fine-tuning paradigm.

\balance

\begin{acks}
This work was supported by the National Natural Science Foundation of China (NSFC) under Grants No.61906180, U21B2046, and Beijing Academy of Artificial Intelligence (BAAI).
\end{acks}

\bibliographystyle{ACM-Reference-Format}
%% \bibliography{sample-base}
\bibliography{my}
\appendix

\end{document}